\documentclass{rspublic}
\usepackage{amsmath}
\usepackage{amsfonts}
\usepackage{amssymb}
\usepackage[Symbol]{upgreek}
\usepackage[nointegrals]{wasysym}
\usepackage{mathrsfs}
\usepackage{graphicx}
\newcommand{\bfu}{{\bf u}}
\newcommand{\bfx}{{\bf x}}
\newcommand{\Upo}{{\Sigma}}
\newcommand{\bdy}{{\partial \Sigma}}
\begin{document}
\title[Planar Navier-Stokes Equations in Bounded Domain]{\large Viscous flow regimes in unit square: Part 4. Vorticity dynamics from monopoles to multipoles}
\author[F. Lam]{F. Lam}
%
%
\label{firstpage}
\maketitle
\begin{abstract}{Planar Navier-Stokes Equations; Vorticity; Stream Function; Non-linearity; Laminar Flow; Transition; Turbulence; Diffusion}

The initial-boundary value problem of the vorticity equation has been solved numerically by an iterative method. A variety of initial vorticity distributions is specified. All of them can be described by simple mathematical functions: there are a vorticity source-sink pair, circular shears out of a localised monopole, a cat's-eye topology and a few flows originating from single or multiple vortices. Our computational results show diverse flow phenomena, such as roll-ups of shear layers, vortex merging or impingement, as well as birth of spiral structures.
\end{abstract}
%
%
\section{Introduction}\label{intro}
The planar Navier-Stokes equations for incompressible flows are  
\begin{equation} \label{ns}
	(\partial_t   - \nu \Delta) \bfu = - (\bfu. \nabla ) \bfu  - \nabla p/\rho, \;\;\; \nabla.\bfu = 0,
\end{equation}
where $\bfu(\bfx)=(u,v)(\bfx)$, and $\bfx=(x,y)$. All symbols have their usual meanings in fluid dynamics. The domain of interest is the square with unit length. The no-slip boundary condition, $\bfu(\bfx){=}0, \; \forall \bfx {\in} \bdy$, applies.
From the continuity equation, we derive the vorticity-stream function formulation  
\begin{equation} \label{vort}
(\partial_t  - \nu \Delta ) \zeta  = -u \partial_x\zeta - v \partial_y \zeta =- \partial_y \psi \; \partial_x \zeta + \partial_x \psi \; \partial_y \zeta, \;\;\;  \Delta \psi =-\nabla {\times} \bfu=- \zeta.
\end{equation}
The initial vorticity is assumed to be bounded and is an arbitrary function of $\bfx$,
\begin{equation} \label{vt-ic}
	\zeta(\bfx,0)=\zeta_0(\bfx), \;\;\; \bfx \in \Upo.
\end{equation}
For $t{\geq}0$, the no-slip implies the boundary stream function satisfies $\psi_{\bdy}=0$. Equivalently, initial velocity $\bfu_0(\bfx)=(\Delta)^{-1}\zeta_0(\bfx)$, where the inverse Laplacian respects the no-slip. If necessary, the pressure is determined, within an arbitrary constant, from (for flows of unit density)
\begin{equation} \label{modp}
p = 2 \: (\Delta)^{-1}\: \big( u_x \: v_y - u_y\: v_x \big), 
\end{equation}
where the inverse is solved subject to Neumann boundary conditions.

The given initial vorticity $\zeta_0$ may not define the self-consistent initial velocity $\bfu_0$ satisfying the no-slip. At the start of computations $t{=}0$, the Poisson equation is first solved to determine $\bfu_0$. At any subsequent time $t{>}0$, the vorticity evolution is governed by the dynamics (\ref{vort}), giving rise to compatible vorticity-velocity fields. 
Table {\ref{runs}} lists the initial configurations tested in the present note.
\section{Vortex configurations}
\begin{table}
	\centering
\begin{tabular}{c|cccc} \hline \hline
Initial configuration (eqn)& $\Delta t \; (\times 10^4)$ & $n$ & $ E_0 $ & $Z_0$ \\ \hline \hline
  Isolated Vortex (\ref{iso}) & $1$ & $1536$ & $0.27$ & $82.13$ \\ \hline
		Exponential (\ref{expn})  		 & $1$ & $512$  & $0.54$ & $97.05$ \\ \hline
		Source-Sink  (\ref{ss}) & $2.5$ & $1024$ & $0.38$ & $389.98$  \\ \hline
		Cat's eye  (\ref{cvt}) & $0.5$ & $2048$ & $0.99$ & $508.18$  \\ \hline
		Hippopede (\ref{hippo})  	& $0.5$ & $2048$  & $1.23$ & $903.17$ \\ \hline 
		Tri-polar (\ref{tripolar1})  	& $2.5$ & $1024$  & $0.18$ & $235.74$ \\ \hline
		Tri-polar (\ref{tripolar3})  	& $2.5$ & $1024$  & $0.06$ & $117.87$ \\ \hline
\hline
				\end{tabular}
\caption{Summary of calculations at fixed viscosity $\nu = 10^{-4}$. The last two columns refer to the energy ($E_0$) and enstrophy ($Z_0$) respectively. Computations at different values of viscosity are attempted. We shall concentrate on the vorticity evolution due to the variety of the initial data. } \label{runs}
\end{table}
\subsection*{Isolated swirl}
An isolated signed vortex may be prescribed by
\begin{equation} \label{iso} 
	\zeta_0(\bfx) = 3\pi \: \frac{\exp ( - \theta^2 \: )}{\sqrt{r}+10^{-3}},
\end{equation}
where the starting vorticity field is everywhere positive, and the parameters are given by
\begin{equation*}
	r = (4x-2)^2+(4y-2)^2 = x_f^2+y_f^2, \;\;\;\;\; \theta = \tan^{-1}(y_f/x_f). 
\end{equation*}
%
\begin{figure}[ht] \centering
  {\includegraphics[keepaspectratio,height=4cm,width=12cm]{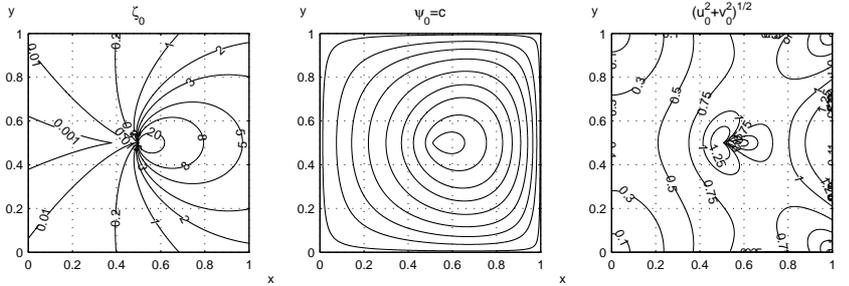}} 
 \caption{Isolated vortex. The data induce noticeable velocities. As soon as the motion is initiated, see figure~\ref{iso10k}, wall layers are generated and the velocities are strongly modified by viscosity so that the no-slip is fully observed. } \label{iso0} 
\end{figure}
\subsection*{Exponential eddy}
\begin{figure}[ht] \centering
  {\includegraphics[keepaspectratio,height=4cm,width=12cm]{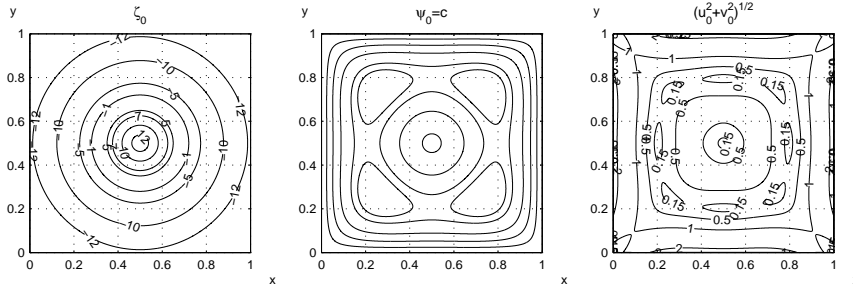}} 
 \caption{Exponential initial vortex (\ref{expn}) at $t=0$. This example is chosen because the vorticity of simple concentric circles induces strong wall velocities while they only produce weak shears near the centre. This is a good case to examine the vorticity production at solid boundaries.} \label{expn0} 
\end{figure}

\begin{figure}[ht] \centering
  {\includegraphics[keepaspectratio,height=12cm,width=12cm]{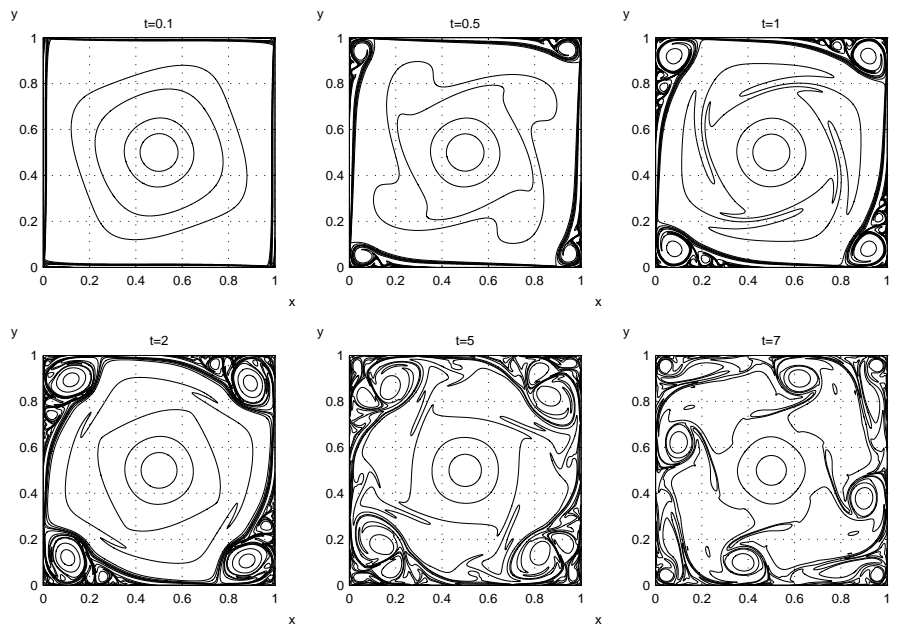}} 
 \caption{Exponential data (\ref{expn}). Plotted iso-contours are $\pm50, \pm25, \pm10$ and $\pm5$. } \label{expn10k} 
\end{figure}

\begin{figure}[ht] \centering
  {\includegraphics[keepaspectratio,height=12cm,width=12cm]{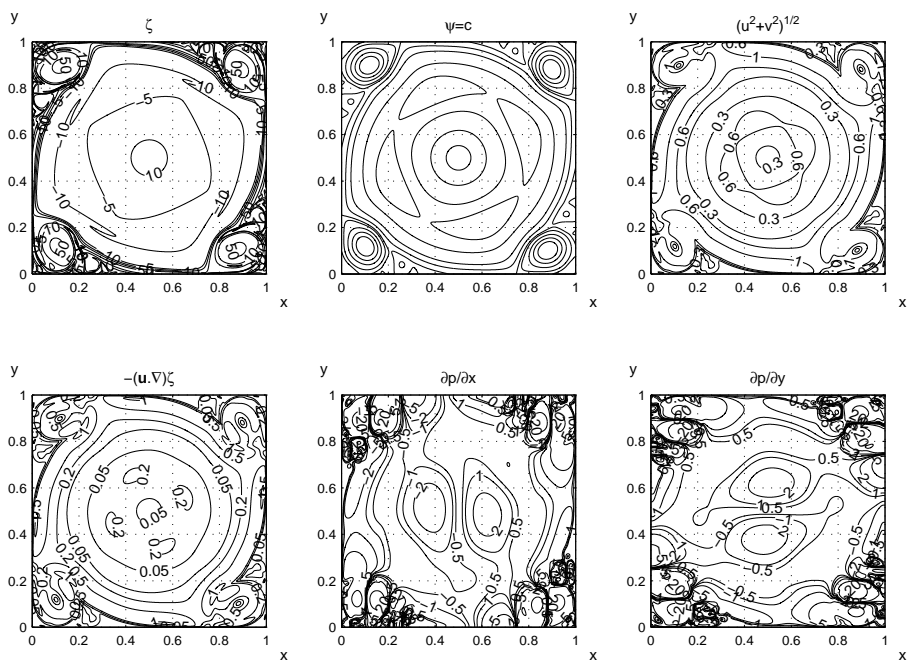}} 
 \caption{Detailed solutions of the exponential vortex at $t=2$.  } \label{expn2p0} 
\end{figure}

The levels of constant vorticity are concentric circles with reduced strength at large radii, see figure~\ref{expn0},
\begin{equation} \label{expn}
\zeta_0(\bfx) = 4\pi \: \big( \; 2 \exp(-r^2) - 1 \;\big).
\end{equation}
%
The computed solutions are displayed in figures~\ref{expn10k} and \ref{expn2p0}.
\subsection*{Source-sink pair}
\begin{figure}[ht] \centering
  {\includegraphics[keepaspectratio,height=4cm,width=12cm]{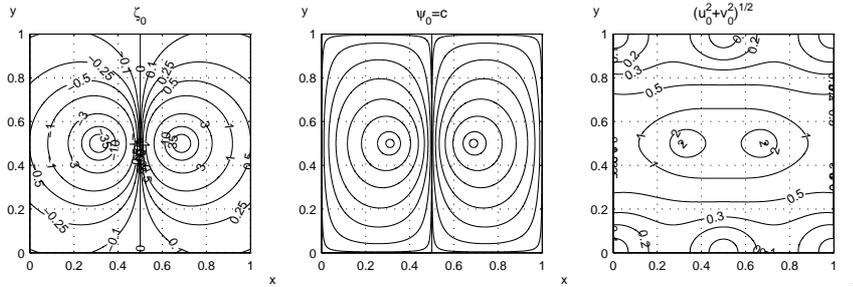}} 
 \caption{Source-sink pair in the square. } \label{sosink} 
\end{figure}

\begin{figure}[ht] \centering
  {\includegraphics[keepaspectratio,height=12cm,width=12cm]{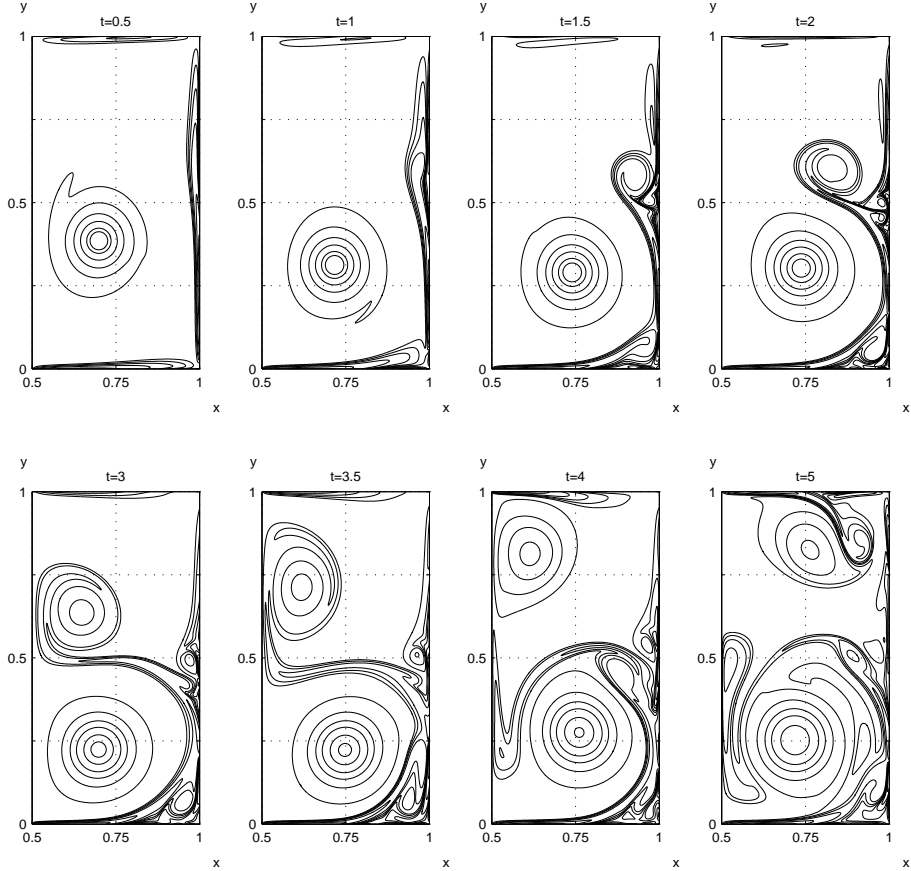}} 
 \caption{Development of the source-sink pair (\ref{ss}). Lines of constant vorticity are drawn at $\pm100, \pm50, \pm20, \pm10, \pm5$ and $\pm2$. The counter-rotating pair induces strong shears on the side walls over the period $0.5<t<1$, thus thickening the wall layers which are then transformed into concentrated eddies by the non-linear effects. Solutions up to $t=5$ show how the wall-generated shears roll-up under mutual interaction. They are further pushed by the source-sink pair until they hit the upper boundary and, subsequently, rebound as two vortex mushrooms. Note that there are extra pairs of wall-shears forming eddies which are squeezed and elongated, after they have become detached from the wall layers. } \label{ss10k} 
\end{figure}

\begin{figure}[ht] \centering
  {\includegraphics[keepaspectratio,height=6.5cm,width=14cm]{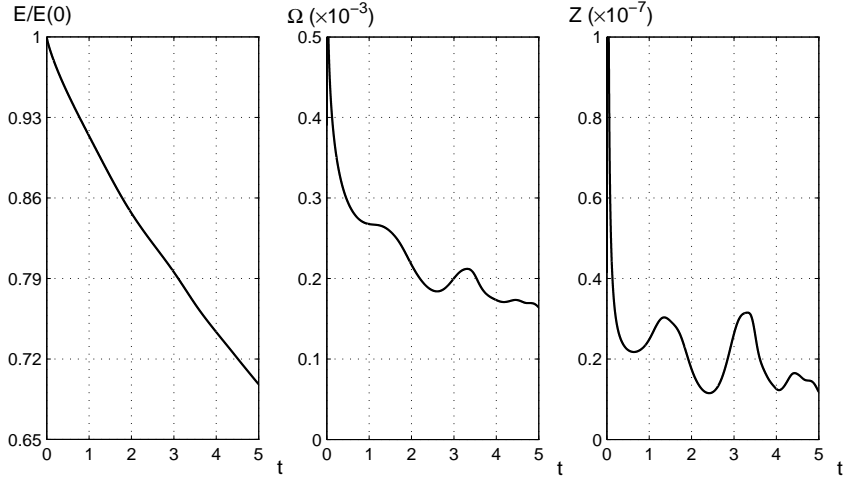}} 
 \caption{The integrated quantities reflect the developing history of the pair in the preceding figure. The oscillations in the enstrophy or palinstrophy are closely related to the birth and re-birth of the small vortices. As a monitoring parameter, we find that the circulation $\Gamma \sim O(10^{-15})\;\forall t > 0$. In addition, mesh convergence is examined at $n=1024$ and $n=1280$.} \label{sshist} 
\end{figure}

A source-sink singularity is a well-known device to model incompressible motions in the theory of potential flow. The topology of the pair suggests strong mutual interaction of the two vortices. We imagine the consequence of a vorticity field having the source-sink data. The initial vortex pair, 
\begin{equation} \label{ss} 
	\zeta_0(\bfx) = \frac{1}{\; 10^{-3}+(x_f{-}3/4)^2+y_f^2 \;} - \frac{1}{\; 10^{-3}+(x_f{+}3/4)^2+y_f^2 \;},
\end{equation}
have the ``regularised singularities'' (figure~\ref{sosink}). The initial velocity setting is a typical doublet where the core speed is no more than $3$. The interaction of the vortices is given in figures~\ref{ss10k} and \ref{sshist}.
\subsection{Cat's-eye: birth of spirals}
\begin{figure}[ht] \centering
  {\includegraphics[keepaspectratio,height=15cm,width=12cm]{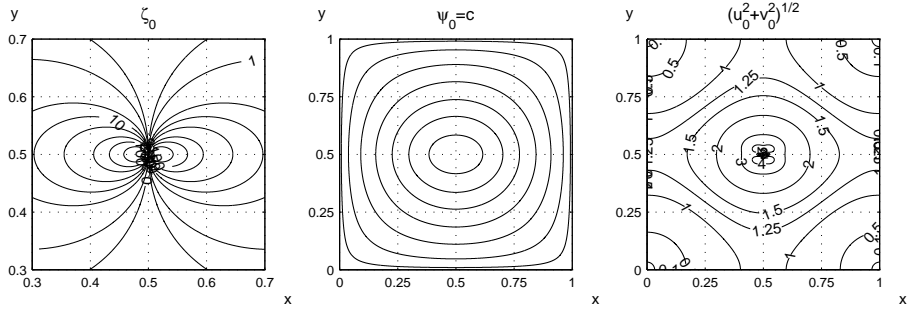}} 
 \caption{Cat's-eye initial data (\ref{cvt}). The core vorticity is stronger and gives rise to a maximum velocity $|\bfu_0| \approx 4$ (incompressible). Because of the high core velocity, the wall layers are ``thickened'' to accommodate the non-slip condition. There is a narrow region next to each wall over which the local velocity is suppressed to zero. Vorticity contours are $0.25,1,5$,$10,20,50$,$100$ and $250$.} \label{cats} 
\end{figure}

\begin{figure}[ht] \centering
  {\includegraphics[keepaspectratio,height=12cm,width=12cm]{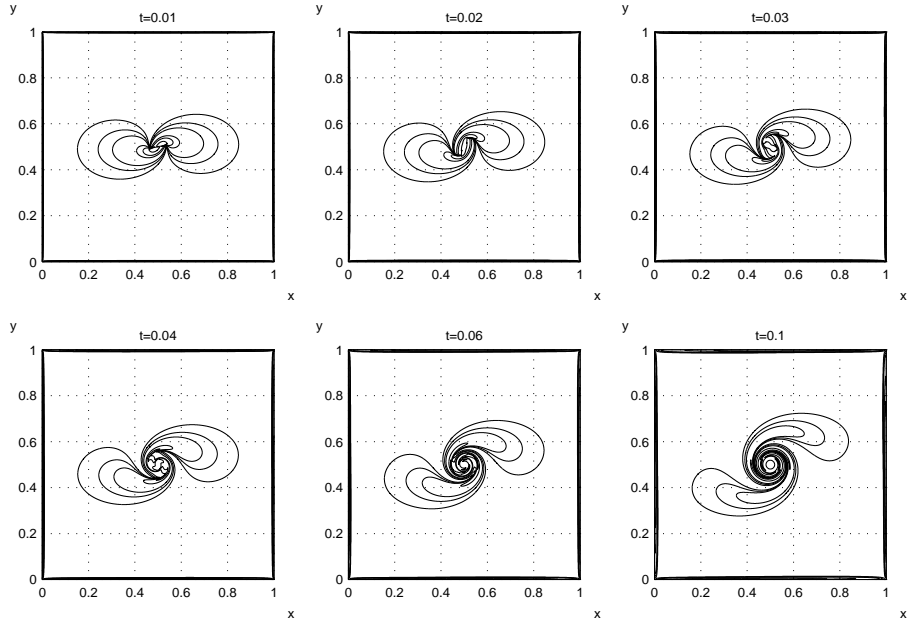}}
 \caption{Cat's-eye vortex evolves over time. The essential point to notice is that our computation has a precise initial condition which cannot be backtracked by extrapolation in time, as we are dealing with a dissipative system. Nevertheless, our example asserts that large-scale spirals revolving a core-body in predominately planar flows are a consequence of the evolution dynamics, which involves no mechanism of flow instabilities. Contours are plotted at $\pm300$, $\pm100, \pm50$, $\pm20, \pm10$, $\pm5$ and $\pm2$.} \label{cats10k} 
\end{figure}

\begin{figure}[ht] \centering
  {\includegraphics[keepaspectratio,height=12cm,width=12cm]{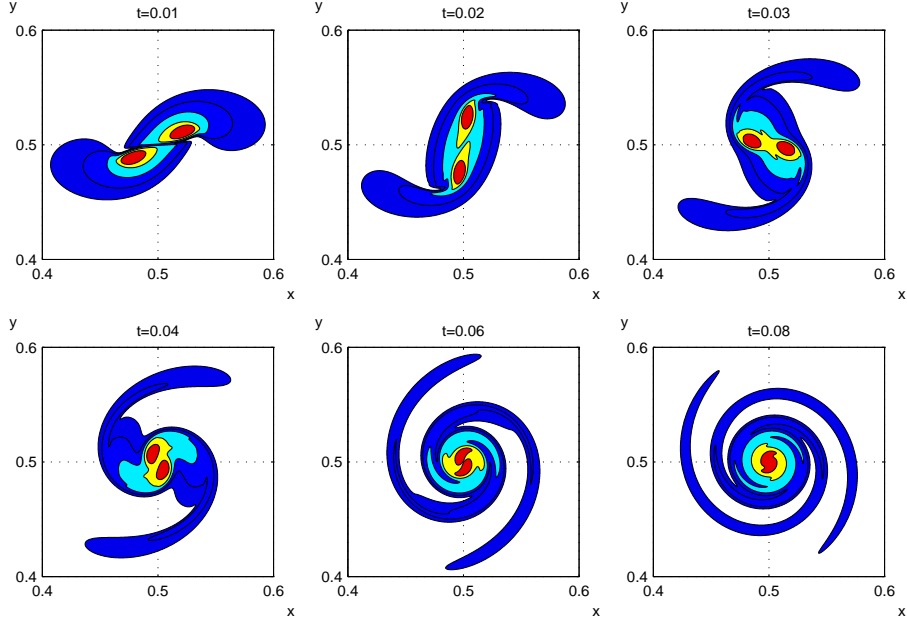}}
 \caption{Zoom-in appearance shows how an S-shaped core has been born out of the revolving eyes during the initial stages of evolution. The outer fan structures are evidently dependent on the core strength and the properties of the fluid. Iso-contour levels are $50,100$, $200,400$ and $600$.} \label{ceyesc} 
\end{figure}

\begin{figure}[ht] \centering
  {\includegraphics[keepaspectratio,height=12cm,width=12cm]{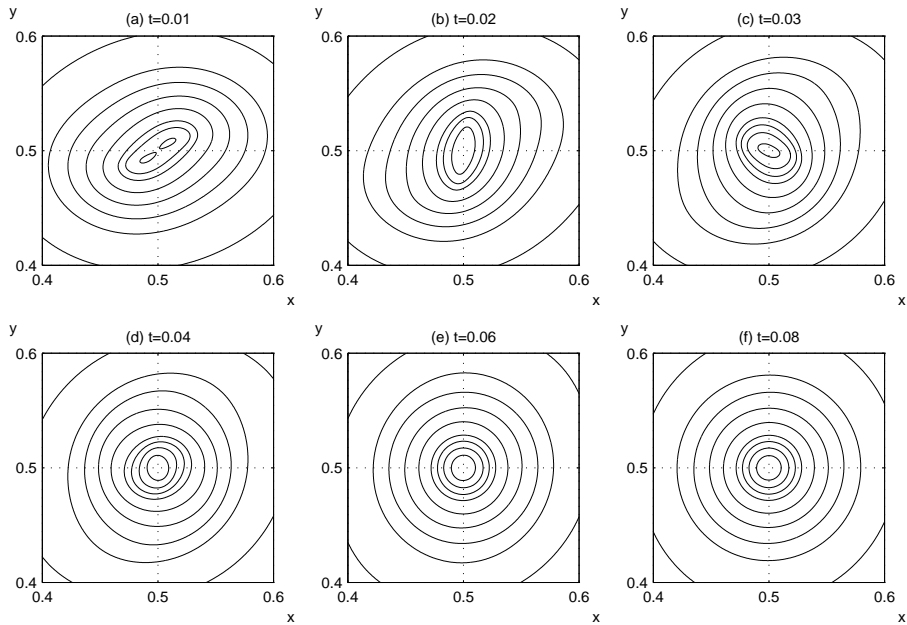}} 
 \caption{Stream function contours $\psi=\const$ It is hard to identify whether the core is in the formation of a spiral structure. Recall that $\psi$ merely plays an auxiliary role.} \label{catspsi} 
\end{figure}

\begin{figure}[ht] \centering
  {\includegraphics[keepaspectratio,height=6cm,width=12cm]{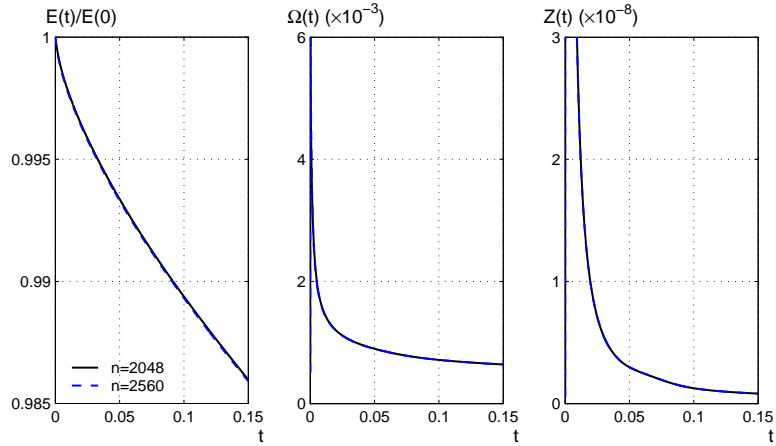}} 
 \caption{Cat's-eye data (\ref{cvt}). The current mesh resolution is satisfactory.} \label{catshist} 
\end{figure}

A cat's-eye initial vorticity can be derived from the theory of potential flow (cf. Art. $156$ of Lamb, 1932). To avoid the singularity, a small quantity, $\delta_c$, is added to regulate the eye. The vorticity is given by
\begin{equation} \label{cvt} 
	\zeta_0(\bfx) = 2\frac{1 +\sin^2x'-\cos^2x'}{\;(\: \cosh y'-\cos x' \:)^2+\delta_c},
\end{equation}
where $x'=4x_f/\pi$, $y'=2y_f$, and the ``singularity'' parameter $\delta_c=10^{-5}$, see figure~\ref{cats}. We shall see that the simple and mild initial data will give rise to swirling shear structures, see figures~\ref{cats10k} to \ref{catshist}.
\subsection*{Hippopede vortex}
\begin{figure}[ht] \centering
  {\includegraphics[keepaspectratio,height=12cm,width=12cm]{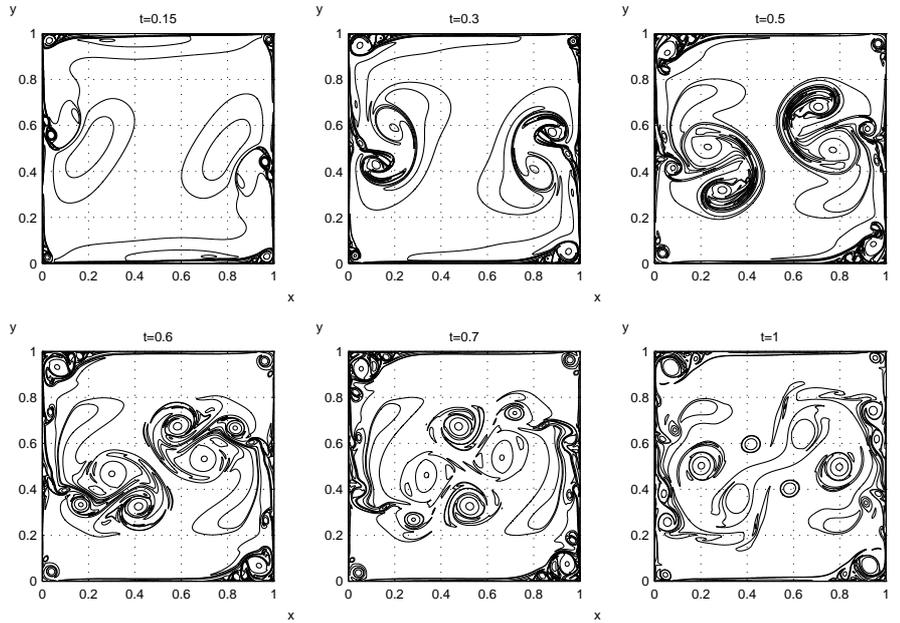}} 
 \caption{Evolution of data (\ref{hippo}) over time. Plotted vorticity contours are $-250,-70$, $-10,-5,20,50$ and $100$. The vorticity has strong disparate gradients on the two side walls. Preliminary test runs indicate that high mesh density is crucial in revealing the fine-scale details, both before and after the collision. The simulation captures the broad features of the head-on impact, cf. the flow visualisation on p.24 of Samimy {\it et al.} (2004).} \label{hippo10k} 
\end{figure}

\begin{figure}[ht] \centering
  {\includegraphics[keepaspectratio,height=6.5cm,width=14cm]{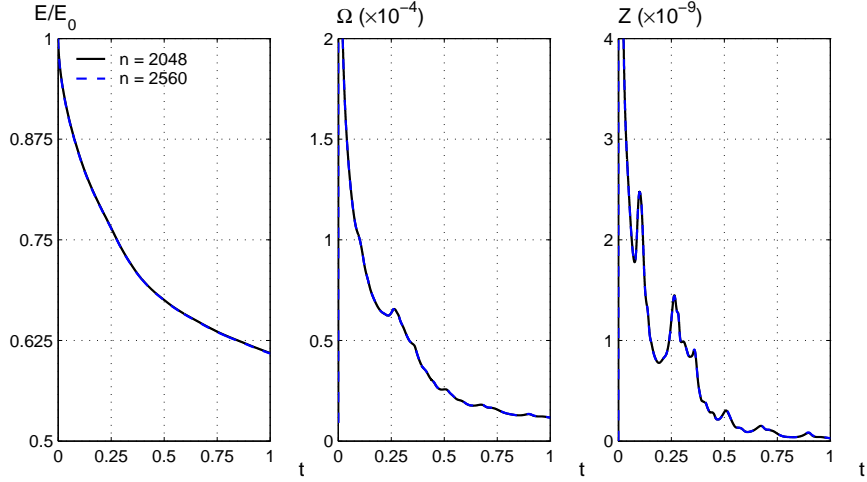}} 
 \caption{This plot confirms the satisfactory mesh convergence for data (\ref{hippo}).} \label{hippohist} 
\end{figure}

This is also called horse fetter (p145 of Lawrence 1972). The zero vorticity curve has a double cusp singularity at the centre. 
\begin{equation} \label{hippo}
\zeta_0(\bfx) = \pi \: \Big( (x_f^2+y_f^2)^2-4x_f^2 \Big).
\end{equation}
%
\begin{figure}[h] \centering
  {\includegraphics[keepaspectratio,height=6cm,width=12cm]{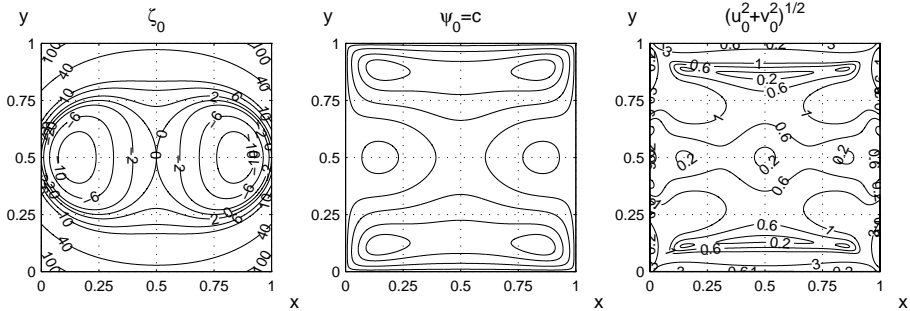}}
 \caption{Hippopede data (\ref{hippo}). The vorticity changes rapidly on the two side walls. } \label{hippo0} 
\end{figure}
The evolution from these data is shown in figures~\ref{hippo10k} and \ref{hippohist}.
\subsection*{Tripolar vortices: shear and smear}
\begin{figure}[ht] \centering
  {\includegraphics[keepaspectratio,height=4cm,width=12cm]{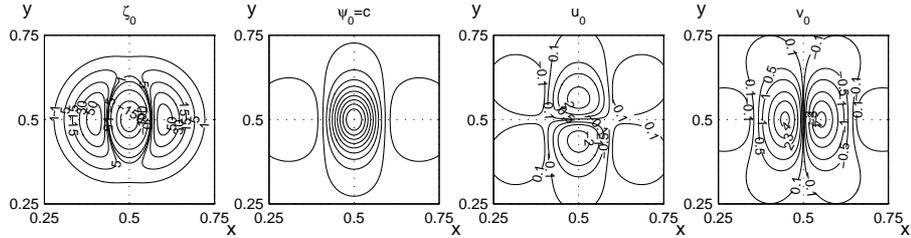}} 
 \caption{The expression (\ref{tripolar1}) defines tri-vortices having a centre core sandwiched by two kidney-shaped vortices. Our simple algebraic definition may be regarded a suitable numerical model to simulate experimental investigations. Note that it is the streamlined rather than the tripolar structure which is conveniently observed in practice. } \label{tripv1} 
\end{figure}

\begin{figure}[ht] \centering
  {\includegraphics[keepaspectratio,height=12cm,width=12cm]{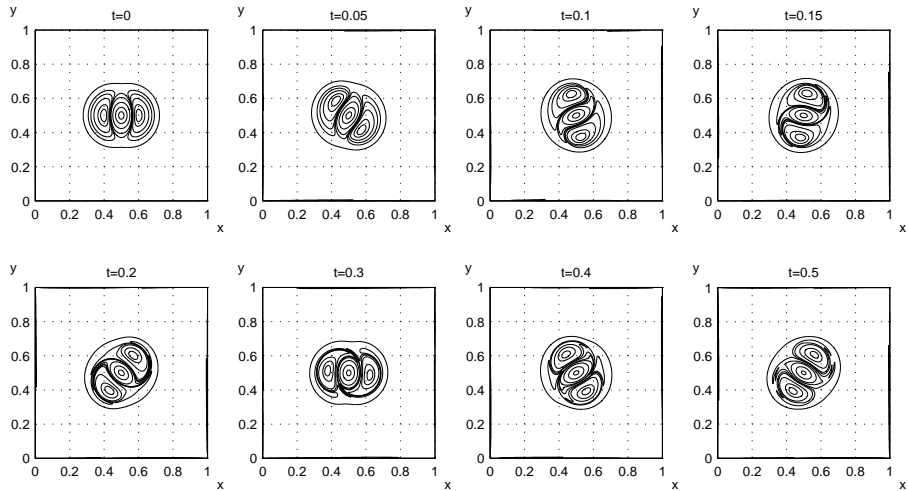}} 
 \caption{Snapshots of the robust tri-polar eddy (\ref{tripolar1}) in rotation. A $360^{\circ}$ turn nearly completes at $t=0.5$ with the sense of the rotation being clockwise, because of the negative stronger central core. Note that the original vorticity at the centre remains largely intact, but is highly smeared while the outer weaker shears are elongated. In the last structure, a mild rotating elliptic core is a double-fan spiral which is expected to decay slowly thereafter. Plotted contours are $-180,-120,-50$, $1,5,20,40,55$ and $\pm10$. The wall viscous layers are thin in the sense that the maximum magnitude of the boundary shears is about $70$ at $t=0.05$, and $12$ at $t=1$. Mesh convergence is investigated at $n=768$ and $n=1024$. } \label{tripv10k1} 
\end{figure}

\begin{figure}[ht] \centering
  {\includegraphics[keepaspectratio,height=8cm,width=10cm]{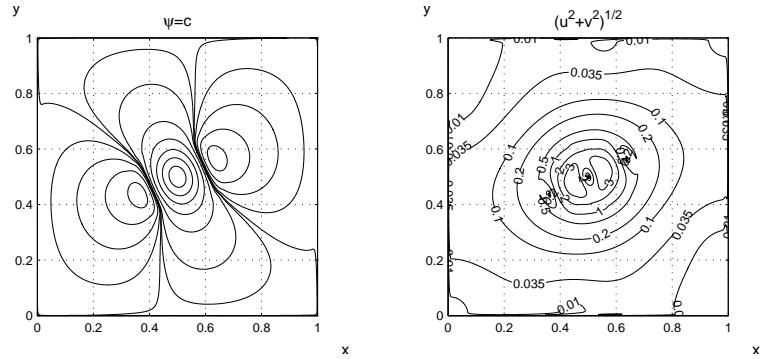}}
 \caption{Stream function and velocity at $t=0.25$ of the preceding plot. They may facilitate the interpretation of visualisation in laboratory conditions (cf. p.15 of Samimy {\it et al.} 2004). } \label{tripv1c} 
\end{figure}

A tri-polar vortex is expressed as 
\begin{equation} \label{tripolar1}
	\zeta_0(\bfx) = 200 \: \exp\big( - 2 \pi^2 (x_f^2+y_f^2) \big) \: \big(16 x_f^2+ 4 y_f^2-1 \big),
\end{equation}
see figure~\ref{tripv1} for details. The particular choice of multiplicative factor $200$ is chosen, from a preliminary study, to make sure the initial vorticity is strong enough to represent interesting dynamics, while the subsequent flow-fields are truly incompressible. 
The calculations are analysed and displayed in figures~\ref{tripv10k1} to \ref{tripv1c}.

\begin{figure}[ht] \centering
  {\includegraphics[keepaspectratio,height=6cm,width=12cm]{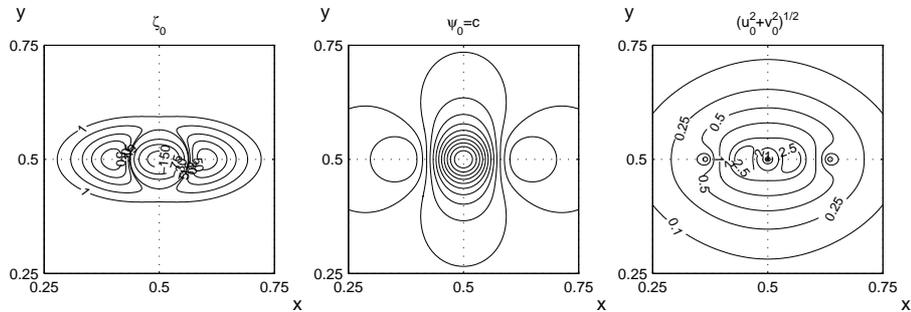}} 
 \caption{The expression (\ref{tripolar3}) defines a tri-vortex with a weighted exponential decay. Its evolution is summarised in figure~\ref{tripv10k3}. } \label{tripv3} 
\end{figure}

An elongated start-up vortex is expressed in 
\begin{equation} \label{tripolar3}
	\zeta_0(\bfx) = 200 \: \exp\big( - 2 \pi^2 (x_f^2+4y_f^2) \big) \: \big(16 x_f^2+ 16 y_f^2-1 \big),
\end{equation}
see figure~\ref{tripv3} for detail.
\section{Remarks}
The examples studied in the present note explain some well-known observations of fluid motion in a quantitative manner. With properly specified initial data, a wide range of problems relevant to applications can be simulated and analysed. The incompressible Navier-Stokes dynamics should have never been restricted to a tiny set of ``steady exact solutions''. 

When a pebble is thrown into a pond, the ripples are in marked difference to those from a cue ball, while a brick will generate a splash. That is why there must be some merits in exploring the exponential or isolated vortex flow. Evidently, every fluid motion is determined by the way it is first generated.

Our computations imply the criticality of generating complex flow fields by vorticity control. Currently, it is fair to say we are pretty poor at it. Indeed, we have not {\it yet} explored practical ways of creating precise vorticity. 

\vspace{1cm}
\begin{acknowledgements}
\noindent 
22 August 2018

\noindent 
\texttt{f.lam11@yahoo.com}
\end{acknowledgements}
\vspace{1cm}
%
\begin{figure}[ht] \centering
  {\includegraphics[keepaspectratio,height=12cm,width=12cm]{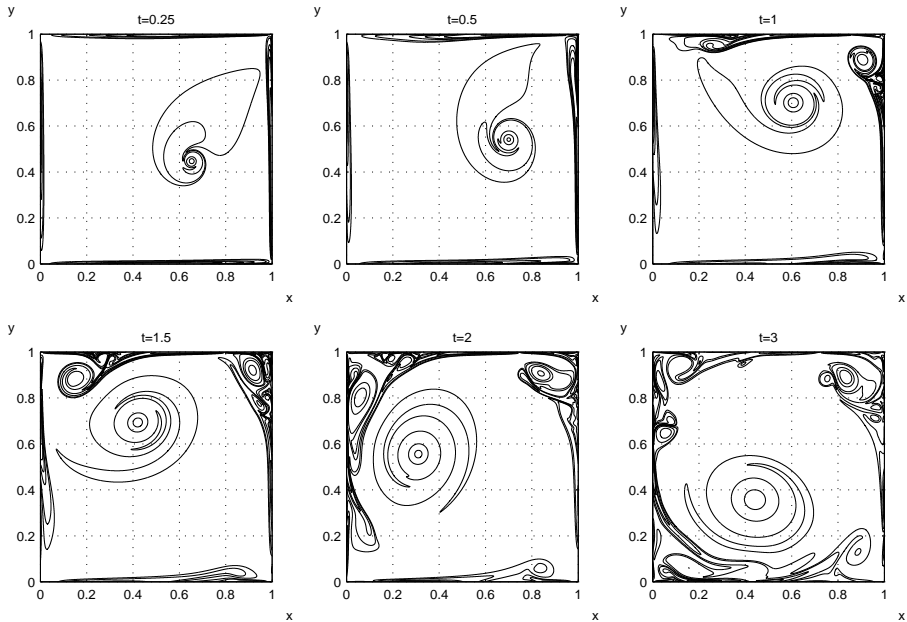}} 
 \caption{Isolated vortex (\ref{iso}) evolves in time. The main vortex is rotating in the anti-clockwise direction. Note that the birth of the wall vortices out of the wall viscous layers. These isolated secondary vortices are the building blocks of the so-called coherent structures in wall-bounded turbulence, even though a ``free-stream'' flow in the sense of boundary layers is absent. The coherence is in short and compressed forms. No difficulties have been encountered during the computations. Iso-contours are $\pm100, \pm50, \pm25$, $\pm10,\pm5$ and $-35$.} \label{iso10k} 
	\end{figure}
%
%
%
%
%
%
%
%
%
%
\begin{figure}[ht] \centering
  {\includegraphics[keepaspectratio,height=12cm,width=12cm]{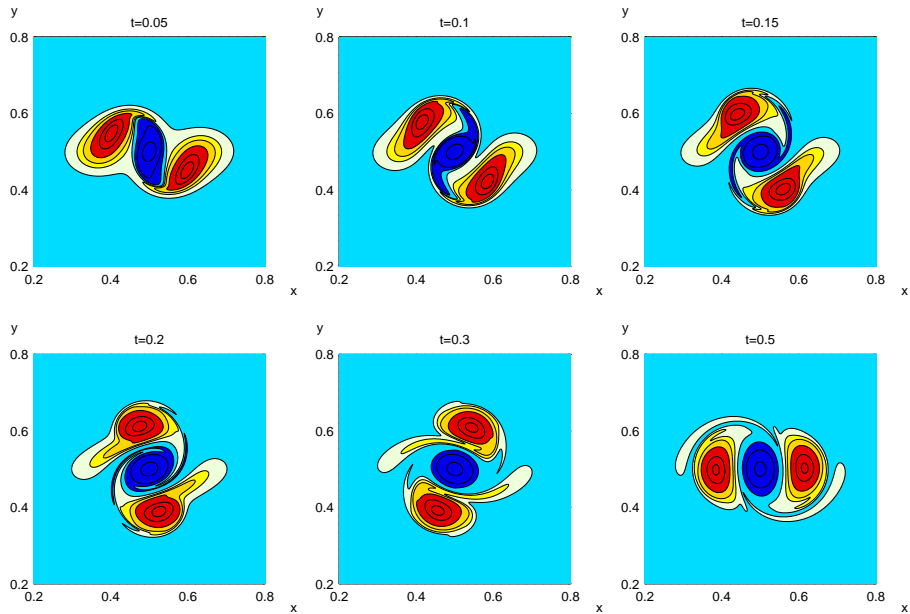}} 
 \caption{Tri-polar vortex (\ref{tripolar3}) in rotation with marked shear patterns over time. Plotted iso-vorticity contours are $-200,-150, -50$, $1,5,20,40,55$ and $\pm10$. } \label{tripv10k3} 
\end{figure}
\end{document}